\documentclass[web]{iosart2c}

\usepackage[numbers]{natbib}
\usepackage{endnotes}
\usepackage{verbatim}
\usepackage{graphicx}
\usepackage{hyperref}
\graphicspath{ {./images/} }

\pubyear{2023}
\volume{0}
\firstpage{1}
\lastpage{1}

\begin{document}

\begin{frontmatter}

\title{Web Intelligence Journal in perspective: an analysis of its two decades trajectory}
\runningtitle{Web Intelligence Journal in perspective}

\author[A]{\fnms{Diogenes Ademir} \snm{Domingos}\thanks{diogenes.20211001317@aluno.uema.br}}
\author[A]{\fnms{Victor Emanuel Santos} \snm{Moura}\thanks{victormoura@aluno.uema.br}}
\author[A]{\fnms{Antonio Fernando Lavareda} \snm{Jacob Junior}\thanks{antoniojunior@professor.uema.br}}
\author[B]{\fnms{Fabio Manoel Franca} \snm{Lobato}\thanks{fabio.lobato@ufopa.edu.br}}
\runningauthor{Diogenes A. D. et al.}
\address[A]{State University of Maranhão (UEMA) \\ São Luís, Maranhão - Brasil.}
\address[B]{Federal University of Western Pará (UFOPA) \\ Santarém, Pará - Brasil.}

\begin{abstract}\label{abstract}
The evolution of a thematic area undergoes various changes of perspective and adopts new theoretical approaches that arise from the interactions of the community and a wide range of social needs. The advent of digital technologies, such as social networks, underlines this factor by spreading knowledge and forging links between different communities. Web intelligence is now on the verge of raising questions that broaden the understanding of how artificial intelligence impacts the Web of People, Data, and Things, among other factors. To the best of our knowledge, there is no study that has conducted a longitudinal analysis of the evolution of this community. Thus, we investigate in this paper how Web intelligence has evolved in the last twenty years by carrying out a literature review and bibliometric analysis. Concerning the impact of this research study, increasing attention is devoted to determining which are the most influential papers in the community by referring to citation networks and discovering the most popular and pressing topics through a co-citation analysis and the keywords co-occurrence. The results obtained can guide the direction of new research projects in the area and update the scope and places of interest found in current trends and the relevant journals.
\end{abstract}


\end{frontmatter}

\section{Introduction}\label{Introduction}

A thematic progression depends on the extent to which studies are synthesized and geared towards detecting areas of conflict and research gaps, since this makes it possible to plan future investigative studies \cite{kumar2020masstige}. Because of this, a critical analysis of a knowledge field can be conducted by employing a systematic literature review; and analyzing position papers, editorials, and other analog artifacts. In particular, a systematic review allows the most significant studies to be determined without any bias or research gaps; this inquiry was undertaken in a strict, replicable, and transparent manner by adopting methodological procedures and reducing the risk of randomness/bias and subjectivity \cite{PAUL2020101717, petersen2011identifying}.

The Web Intelligence Consortium (WIC) is an international organization devoted to encouraging world scientific research in Computational Web Intelligence (WI), at research centers and among WI members throughout the world, through collaborations, conferences, workshops, and journals \cite{WebIntelligence}.  WI research - in its form as the use of Artificial Intelligence (AI) - impacts thematic areas such as the Web of People, Web of Data, and Web of Things \cite{WebIntelligence}. 

The Web Intelligence Journal (WIJ) publishes four editions a year (IOS Press), is peer-reviewed, and includes research papers of a high standard in every area that belongs to the WI field. Its primary focus is on the following areas: Web of People, Web of Data, Web of Things, Web of Trust, Web of Agents, and Emerging Web Technology for Healthcare and Intelligence in the 5G era \cite{WebIntelligence}. Furthermore, the review has an h5-index equal to 24 and an h-5 median equal to 38, by the Google Scholar Metrics \footnote{available at  \url{https://shre.ink/1gTZ}, accessed on  04/09/2022.} and Scimago Journal $\&$ Country Rank \footnote{available at  \url{https://shre.ink/1ZyZ}, accessed on 04/09/2022}. In addition, the journal had a Citation impact of 1.8 on the CiteScore between 2018 and 2021. This metric is based on the number of citations of documents published in four years, divided by the number published in the same four years that were indexed in the Scopus database.

Because the WIJ had a 20th Edition which was aimed at providing an in-depth review of its research fields and a visionary exploration of its prospects, an opportunity arose to carry out a study that could examine how this community evolved through an analysis of research teams, individual productivity, and the related networks. The study can be distinguished from other SR projects because it does not seek to concentrate on a particular subject. We provide an overview of the publications that have appeared in the WIJ over 20 years and thus show the main papers and topics that have emerged from the literature. The study employs rigorous analytical methods and synthesizing procedures to achieve this and relies on bibliometric techniques to create a quantitative map of the WIJ publications \cite{ellegaard2015bibliometric}. 

As well as this, the bibliometric mapping makes it possible to i)assess the academic links between the publications, ii) highlight the key topics, iii) visualize the impact of the papers, iv) determine the nature of current research, and v) lay down guidelines for future projects \cite{pritchard1969equity, jones2014future}.  Given this, we are seeking to answer the following research questions through a bibliometric analysis of the works published in the area: 

\begin{enumerate}
    \item RQ1: How has the WIJ community evolved so far - i.e. up to the second decade of its existence?
  \item RQ2: What are its most influentialworks?
  \item RQ3: What are the current trends regarding the publications and most popular topics emerging from the WIJ community?
\end{enumerate}

The results obtained in this study show how the WIJ Community has evolved. Some of the analyses that were conducted can assist the WIC in planning research programs and formulating strategies. This can lead to sustainable growth by mapping out the research fields, ensuring the less explored are examined in greater depth, assessing future prospects, and embarking on a possible review of the scope and key topics of the journals and related events that have taken place.

The remainder of the article is structured as follows. The materials and methods employed are outlined and discussed in Section  \ref{materials}. The results are presented and discussed in Section \ref{results}. Finally, the conclusion, threats to the research validity, and suggestions for future work can be found in Section  \ref{FinalConsiderations}.

\section{Materials and Methods}\label{materials}

A systematic review of the literature was carried out to provide a detailed account of the current state of research in WIJ and to make a quantitative assessment of the related literature. This was based on the    Scopus database of Elsevier, and corroborated with bibliometric mapping. According to \cite{pritchard1969equity}, the mapping enables a statistical mapping to be made of the scholastic relationships between the published works, with a view to outlining the most important topics published by the WIJ. This is undertaken by referring to  the papers that are most cited, as well as by showing the trends of current research and the guidelines laid down for future work \cite{jones2014future}.

A systematic literature review was carried out to provide a detailed account of the current state of research in WIJ and to quantitatively assess the related literature. This was based on the    Scopus database of Elsevier, and corroborated with bibliometric mapping. According to \cite{pritchard1969equity}, the mapping enables a statistical mapping the  relationships between the published works to outline the most critical topics published by the WIJ. This is undertaken by referring to the papers that are most cited, as well as by showing current research trends and the guidelines laid down for future work \cite{jones2014future}.

When faced with the vast amount of scientific material available, the state-of-the-art includes techniques that can allow a review of the literature to be undertaken that has benefits such as the ability to form flexible databases for the papers, which can thus be easily searched and upgraded \cite{pickering2014benefits}. Several classical approaches can be mentioned, such as the Knowledge  Development  Process –Constructivist (Proknow-C) \cite{ProKnow-C}, which follows four stages: selection of the bibliographic portfolio; a systematic analysis of the portfolio; planning of research objectives; and the  Methodi Ordinatio methodology \cite{MethodiOrdinatio}, which makes use of Information and Communications Technology (ICT) to assist in the collection, selection, and classification of scientific papers. This is based on an assessment of scientific significance, taking into account three factors: degree of impact, year of publication, and the number of citations.

\begin{figure}[!htb] 
  \includegraphics[width=.8\columnwidth]{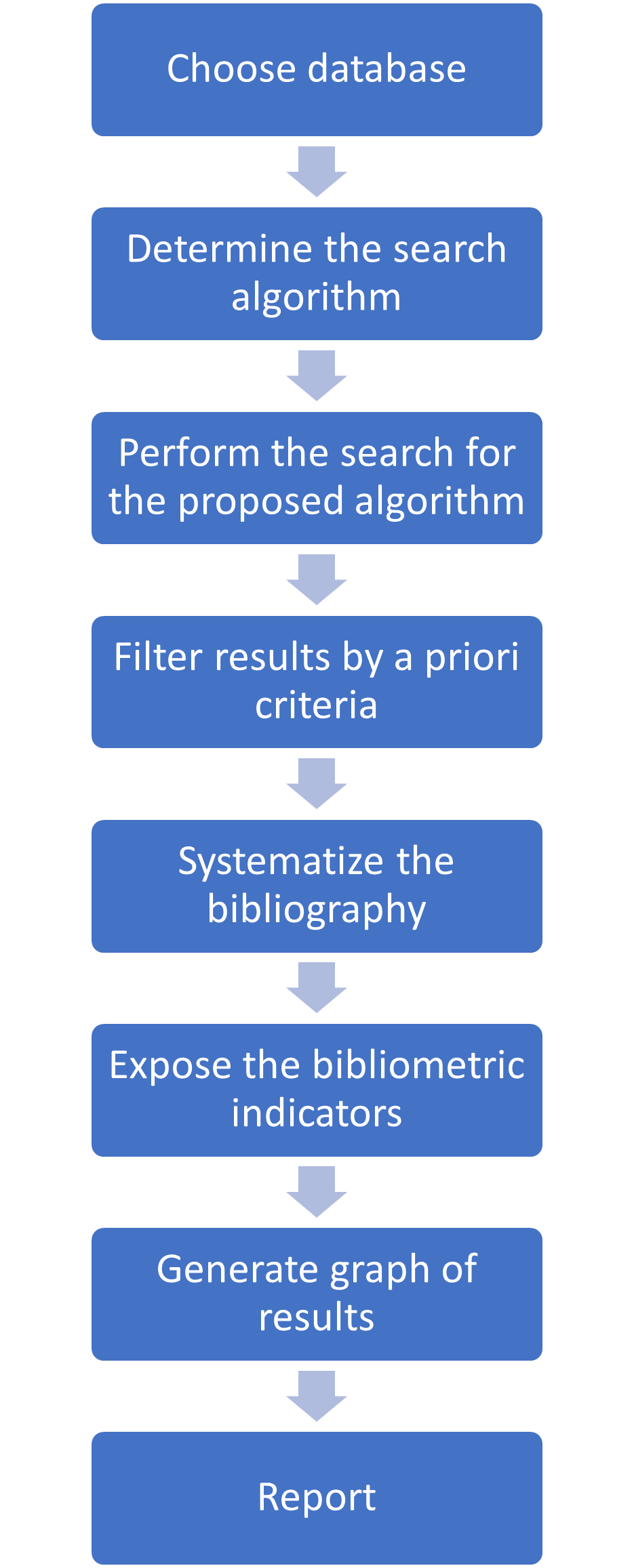}
  \caption{Steps taken in the literature review based on \cite{kitchenham2004procedures}.}
  \label{fluxo}
\end{figure}

However, unlike traditional systematic reviews, our approach will only take account of specific stages in the general SR process because it aims to analyze the whole set of papers published by WIJ rather than concentrating on one particular subject. Thus, some stages of the SR will not be carried out, such as the alignment to the subject and filtering data based on a list of keywords employed by Proknow-C; in addition, the Methodi Ordinatio methodology will not be applied. The general process \cite{kitchenham2004procedures} of a literary review is illustrated in Fig. \ref{fluxo}.

By following the general procedure, the data used in this work were collected from the Scopus database. The Scopus database was chosen instead of others, such as Web of Science (WoS), because it had extensive content, presenting a trade-off between coverage and data cleansing \cite{wang2016large, zhao2015analysis}. Besides, it was decided not to use Google Scholar because this index has less output in English, and the number of its citations has a solid statistical dependence on the citations in  WoS/Scopus, according to the  Spearman correlation coefficient  \cite{martin2018google}.

In this way, it could be determined that the data collected were from publications attached to the  International Standard Serial Number (ISSN and E-ISSN): 1570-1263, 1875-9289 (online), with publications between  2003 and 2016; and 2405-6456, 2405-6464 (online) with publications from   2015 until today, since this criterion met the research objectives. The only documents selected were from the type of paper and excluded editorials, reviews, and conference works. Works published up to 26th September, 2022 were included, and only papers in English were kept.

Stages of sanitization were required in the data pre-processing such as for example normalization, cleansing, standardization etc., similar to the work \cite{lobato2020social}. It was necessary to carry out a systematized management of the bibliography, given the objectives of the research. The following stages in the general process (indicators, graphs and reports) will be discussed in Section \ref{results}.

Stages of sanitization were required in the data pre-processing, such as normalization, cleansing, standardization, etc., similar to the work \cite{lobato2020social}. It was necessary to carry out a systematized bibliography management, bearing in mind the research objectives. The following stages in the general process (indicators, graphs and reports) will be discussed in Section \ref{results}.

In view of the multitude of analyses, the present work has been restricted to specific tasks so that the focus can be kept on the research questions. Thus, an analysis was conducted on the number of publications and citations to characterize the community. In light of this, there were analyses of co-citations and co-occurrences to determine the existence of inter-institutional collaborations and assess the impact of the publications. Finally, a study of the research themes was conducted by creating a bibliometric mapping of the framework of scientific fields using the  VOSviewer software tool 1.6.18 \cite{van2010software} and Cytoscape \cite{Cytoscape}, widely adopted in the literature. The scripts for the analyses were written in the Python 3.8.16 language, with the aid of the pandas, numpy, and matplotlib libraries. For replicability, the scripts and data that were collected were arranged in the GitHub of the authors \footnote{Scripts available in  \url{https://github.com/suppressed-blind-review}}.

\section{Results and discussion}\label{results}

This section presents and discusses the obtained results. They are grouped according to the research questions.

\subsection{How did the  WIJ Community evolve up to the second decade of its existence?} \label{RQ1}

As mentioned, we analyzed the publications between 2003 and 26th September 2022. Table \ref{tab_pub_citations_por_ano} illustrates the number of published papers and citations per year. It should be noted that 411 papers were published in these 20 years. The year with the most significant number of publications was 2012, with 25 works, followed by 22 and 21 papers in 2014 and 2013, respectively. 

\begin{table}[!htb]
\caption{Number of Publications, citations, and authors by year.}
\small
\renewcommand{\arraystretch}{1}
\renewcommand{\tabcolsep}{20 pt}
\label{tab_pub_citations_por_ano}
\centering
\resizebox{\columnwidth}{!}{%
\begin{tabular}{crrr}
\hline
\textbf{Year} & \multicolumn{1}{c}{\textbf{Publications}} & \multicolumn{1}{c}{\textbf{Cited}} & \multicolumn{1}{c}{\textbf{Authors}} \\ \hline
2003                                & 16           & 539           & 44            \\
2004                                & 12           & 99            & 31            \\
2005                                & 16           & 131           & 32            \\
2006                                & 22           & 417           & 67            \\
2007                                & 25           & 258           & 68            \\
2008                                & 24           & 315           & 59            \\
2009                                & 24           & 287           & 69            \\
2010                                & 24           & 167           & 73            \\
2011                                & 24           & 171           & 57            \\
2012                                & 25           & 133           & 71            \\
2013                                & 21           & 63            & 62            \\
2014                                & 22           & 92            & 58            \\
2015                                & 15           & 112           & 39            \\
2016                                & 20           & 147           & 52            \\
2017                                & 17           & 28            & 42            \\
2018                                & 20           & 58            & 73            \\
2019                                & 29           & 127           & 76            \\
2020                                & 19           & 18            & 42            \\
2021                                & 24           & 1             & 57            \\
2022                                & 12           & 0             & 29            \\ \hline
\multicolumn{1}{l}{\textbf{Avg}}    & 21           & 158           & 55            \\
\multicolumn{1}{l}{\textbf{Median}} & 22           & 129           & 58            \\
\multicolumn{1}{l}{\textbf{Std}}    & 4.64         & 141.66        & 15.29         \\
\multicolumn{1}{l}{\textbf{Total}}  & 411          & 3163          & 1101          \\ \hline
\end{tabular}%
}
\end{table}

The analysis showed that the average number of published WIJ papers in these 20 years is approximately 21, with a median of 22 and a standard deviation of 4.64. It is not just a question of assessing the evolution of the community in terms of the number of published papers but also entails the growth in the number of accompanying researchers. For this reason, our study also analyzed the involvement of the researchers in the community. The results showed that there were 1,101 WIJ authors in this period of twenty years. Concerning this figure, the first peak was detected in 2010, when there were 73 authors of published papers. The same peak appeared again in 2018. However, the summit was reached in 2019 with 76 authors. It should be noted that the average number of authors per year was about 55, with a median of 58 and a standard deviation of 15.29. Document sanitization was required during the analysis so that homonyms could identify the authors; for example, five authors were found with the name Li X.

Moreover, the analysis showed that 41 authors had submitted papers solely written by themselves, meaning there was only one paper per author. It was also found that, on average, a collaboration involving three authors per published paper. In addition, the maximum number of authors found to be responsible for a single paper was 10. This was the case with the paper entitled ``COBRA - Mining web for corporate brand and reputation analysis'' \cite{spangler2009cobra}, which is concerned with finding a solution (called COBRA) for image detection and the reputation of a trademark. This goes beyond researching keywords, which often leads to an excessive amount of information being handled manually, and involves employing text data mining techniques on the Web to obtain knowledge of both structured and non-structured sources.

\begin{figure*}[!htb] 
  \includegraphics[width=.7\linewidth]{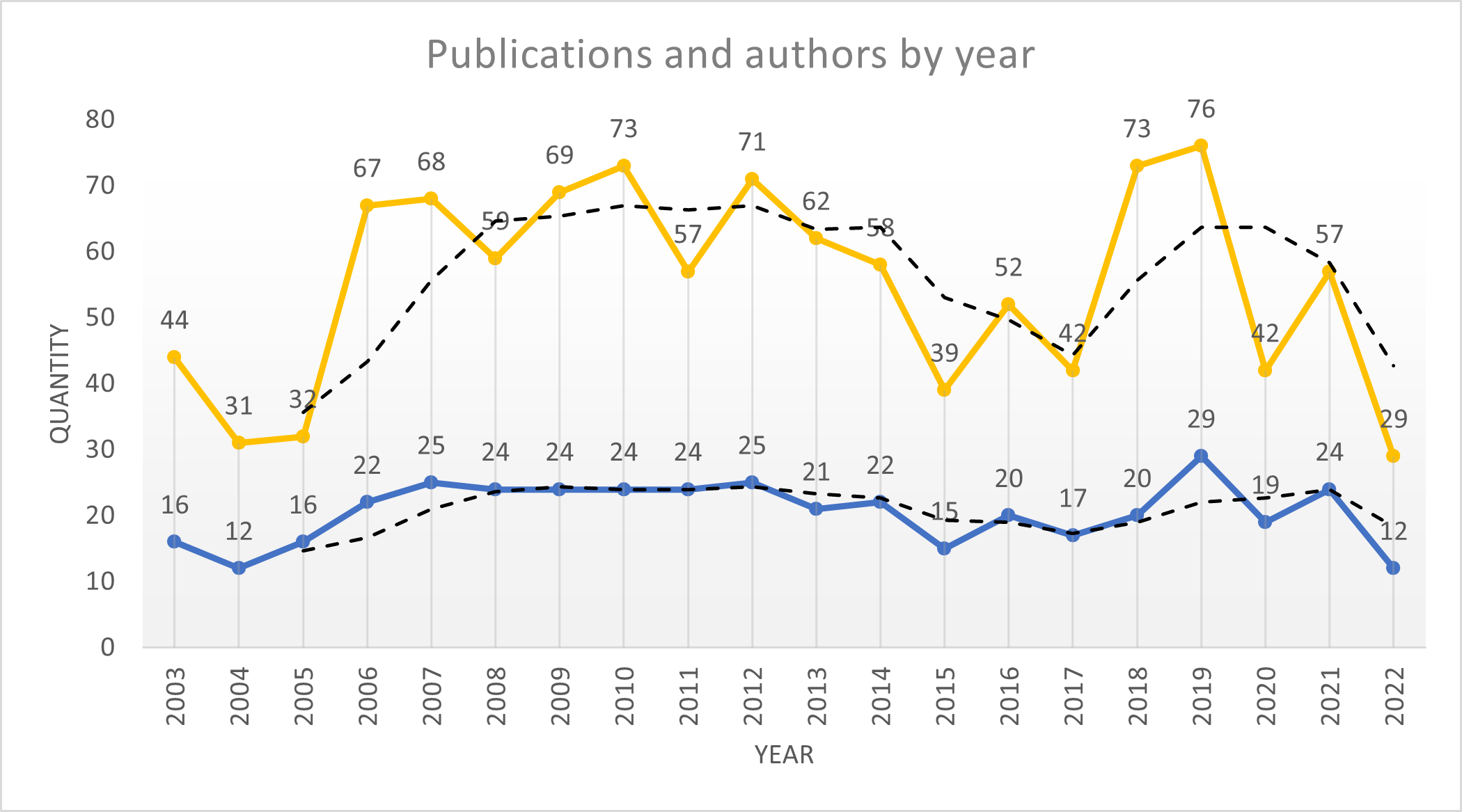}
  \caption{Publications by year (2003–2022*).}
  \label{fig_pub_e_autor_por_ano}
\end{figure*}

\begin{figure*}[!htp] 
  \includegraphics[width=.7\linewidth]{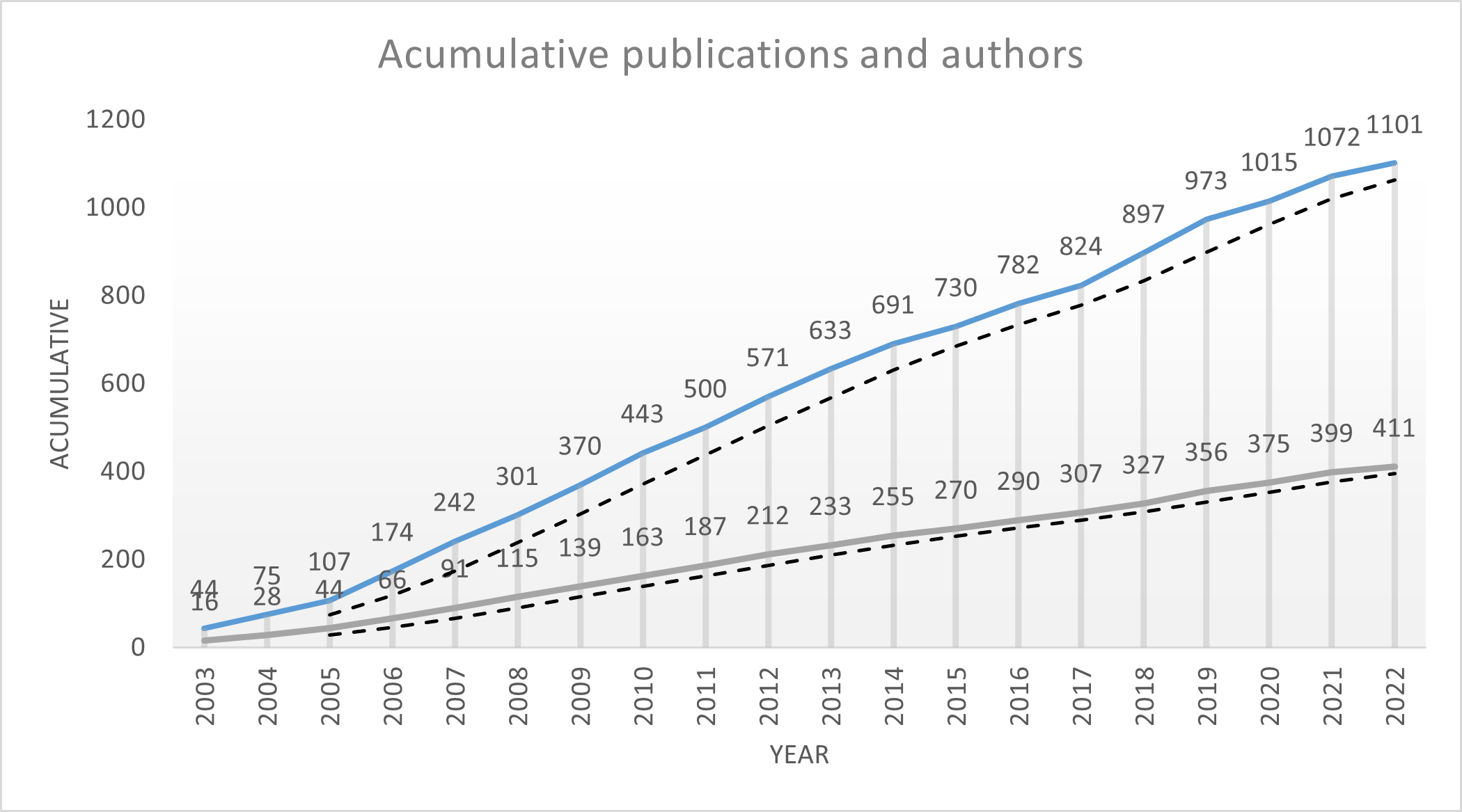}
  \caption{Cumulative publications and authors (2003–2022*).}
  \label{fig_cumulative_pub}
\end{figure*}

In light of this, the data are displayed as graphs. Fig. \ref{fig_pub_e_autor_por_ano} shows the moving average (dashed line) of the number of authors and publications per year, while Fig. \ref{fig_cumulative_pub} shows the cumulative average growth rate of the different publications and authors from 2003 to 26/09/2022. The blue line indicates the number of authors, and the grey line indicates the number of publications per year. The dashed line shows the moving average based on the last three years. Thus the growth of one community compared with another in the last 20 years is clearly apparent - the cumulative growth of publications and the number of new authors in each edition, where an evolving pattern, is evident.

An analysis made it possible to determine which authors had the most significant number of publications in these two decades.   Table \ref{top_autor} the top 10 authors attached to the published papers. It can be seen that an author named Tao X. from Queensland University of Technology, Australia, had seven publications between 2007 and 2021, and his papers had 72 citations. Another author, Treur J., affiliated with Vrije Universiteit Amsterdam in the Netherlands, had six publications between 2009 and 2013 with 40 citations altogether. Bosse T., also affiliated with Vrije Universiteit, published five papers that were cited 40 times. Ursino D., affiliated with Università Mediterranea di Reggio Calabria, Italy, had five publications between 2004 and 2012, which obtained 20 citations in this period. Li L. from the Wuhan University of Technology, China, was linked to five publications, although only one of them was cited five times. Vercouter L, affiliated with INSA in Rouen, France, had four publications between 2012 and 2017 with 19 citations altogether. Liu L., affiliated with Chongqing University, China, published three papers in 2018 alone, which had eight citations. Still, as in the case of  Li L., only one of the papers was cited. Lesser V., affiliated with the University of Massachusetts Amherst, United States, published three papers between 2004 and 2013 that had 23 citations altogether. The penultimate author, Fan X., affiliated with Pennsylvania State University, United States, had three publications between 2004 and 2014, which obtained 12 citations. Finally, the last author of the top 10, Yu Y., affiliated with Shanghai Jiao Tong University, China, published three papers between 2006 and 2012 which obtained ten citations.

\begin{table}[!htb]
\centering
\label{top_autor}
\caption{Top 10 authors.}
\renewcommand{\tabcolsep}{20 pt}
\resizebox{\linewidth}{!}{%
\begin{tabular}{rlrr}
\hline
\multicolumn{1}{c}{\textbf{Rank}} & \multicolumn{1}{c}{\textbf{Author}} & \multicolumn{1}{c}{\textbf{Documents}} & \textbf{Cited by} \\ \hline
1  & Tao X.       & 7 & 72 \\
2  & Treur J.     & 6 & 40 \\
3  & Bosse T.     & 5 & 40 \\
4  & Ursino D.    & 5 & 20 \\
5  & Li L.        & 5 & 5  \\
6  & Vercouter L. & 4 & 19 \\
7  & Liu L.       & 4 & 8  \\
8  & Lesser V.    & 3 & 23 \\
9  & Fan X.       & 3 & 12 \\
10 & Yu Y.        & 3 & 10 \\ \hline
\end{tabular}%
}
\end{table}

Owing to the techno-scientific character of the study, it was found that the community members were attached to some institutions. The work carried out included an analysis of this academic network. Even though the data obtained by Scopus contained the institutional affiliation description of each author, some pre-processing steps had to be planned so that the names of the institutions could be standardized. This meant that soon afterward, an affiliate network could be created, and it was possible to make an assessment of these connections/links.

\begin{figure}[!htb] 
  \includegraphics[width=1\columnwidth]{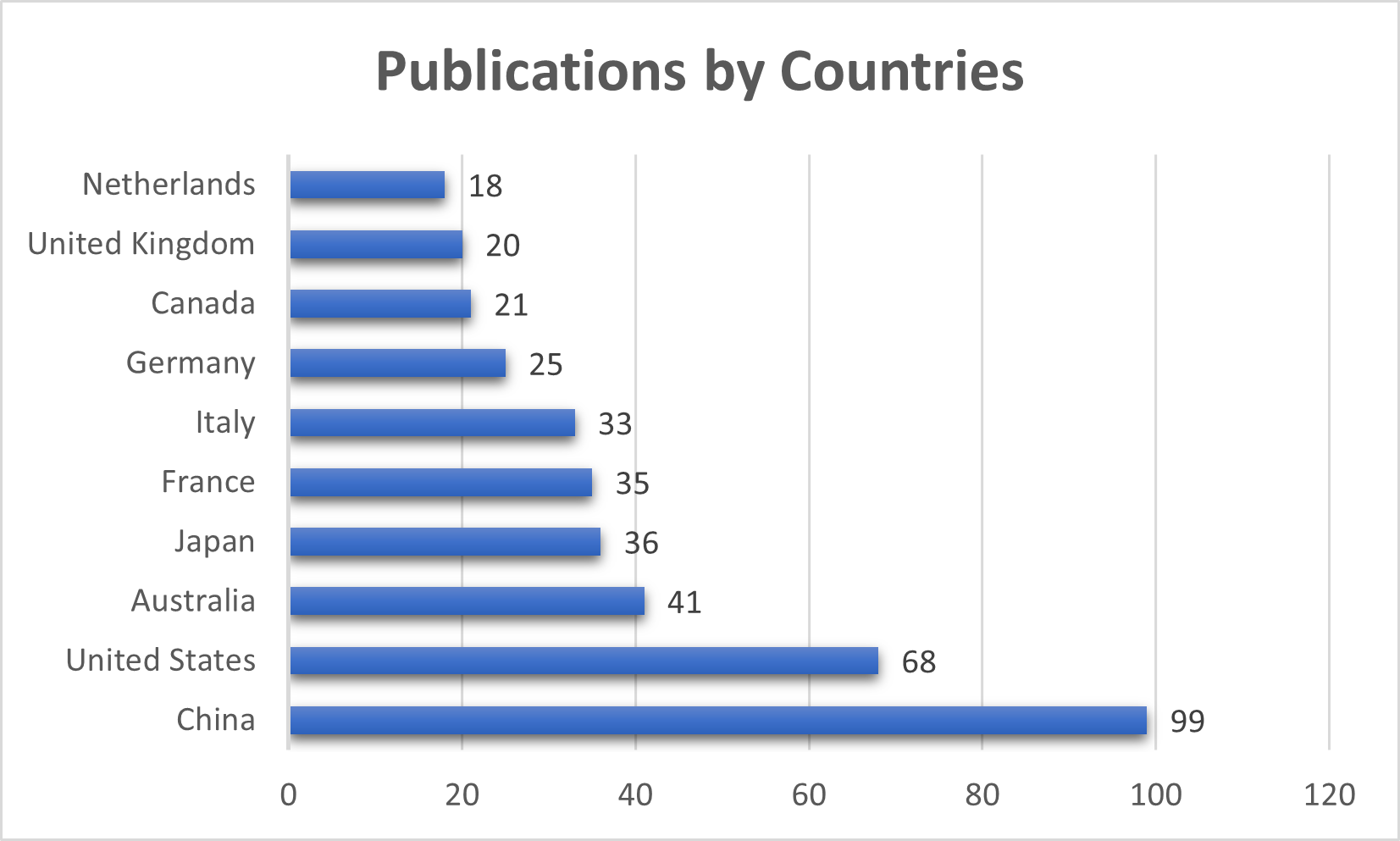}
  \caption{Publications by countries (2003–2022*).}
  \label{top_10_countries}
\end{figure}

\begin{figure*}[!htb] 
  \includegraphics[width=1\linewidth]{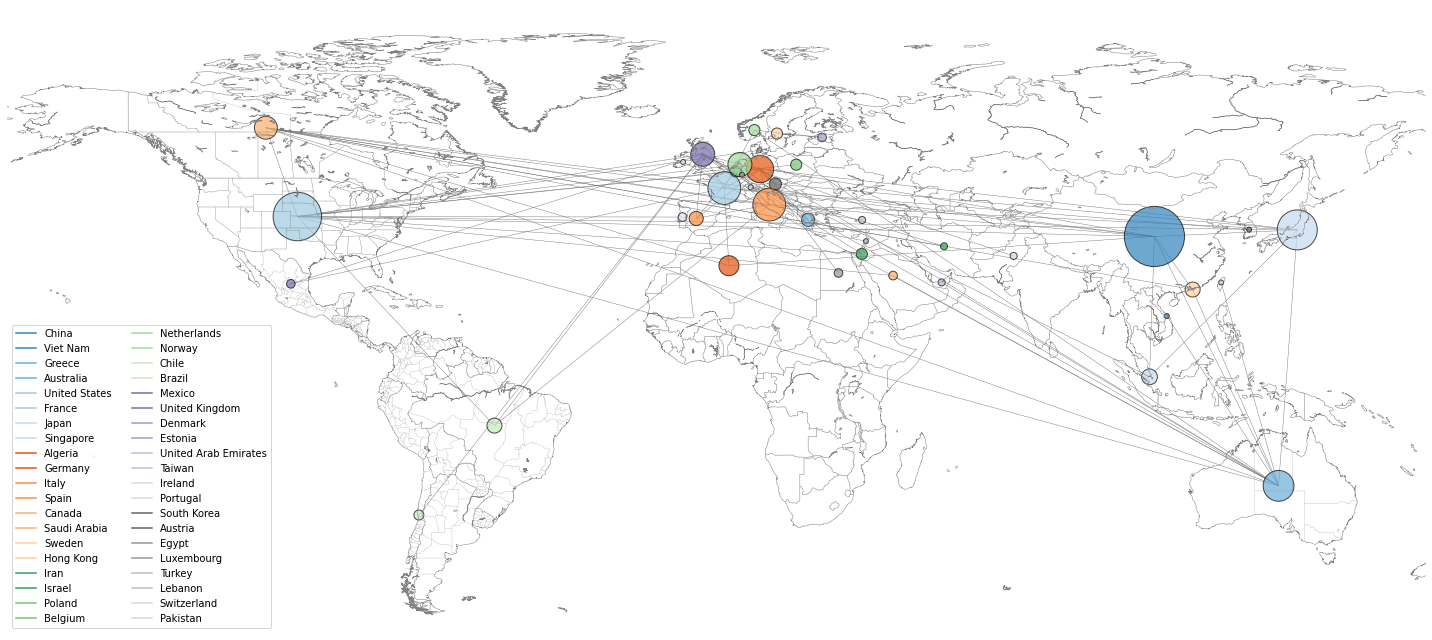}
  \caption{Geographical distribution of institutions.}
  \label{mapa}
\end{figure*}

The results show that WIJ had global coverage. Our analysis demonstrates that during the two decades, academic publications were attached to a wide range of institutions worldwide. Fig. \ref{top_10_countries} shows the ten countries with the most significant number of publications. In contrast, Fig. \ref{mapa} provides a geographical distribution of the institutions weighted by the number of published papers. 

According to Fig. \ref{top_10_countries}, the institutions in China had the highest number of WIJ-published papers, with 99, followed by the institutions in the US with 68 documents, and in the 3rd place, the institutions in Australia with 41. At the other end of the scale, the organizations in the Netherlands published 18 documents, the UK (20) and Canada (21). At this point, it is worth mentioning that the institutions based in Brazil did not reach the top 10; however, they played an active role in the community and were in 16th place with ten papers published during the two decades of the Journal's existence. Concerning this, Table \ref{top_instituicoes} is displayed to name the institutions that were most active in the community throughout this time. The research found that Vrije Universiteit in Holland was the university that most frequently had publications, with ten papers. The University of Tokyo, Japan, followed in 2nd place with eight publications. It should be noted that the Chinese Academy of Science, China, appeared in 3rd place, even though it had the most significant number of published works. At the end of the ranking were the Ministry of Education, China, and the University of Calgary, Canada, with five publications each during these two decades.

\subsection{What are the most influential papers?} \label{RQ2}

We analyzed the citation networks of the papers, as quantified in Table \ref{tab_pub_citations_por_ano}. Our analysis showed that the 411 works published in these 20 years of WIJ yielded 3,163 citations, with an average of 158 per year. In addition, it was found that the papers with the largest number of citations were those in 2003. A trend showed that the greater the number of papers published, the larger the number of citations. Only the papers published in 2003 yielded 539 citations. Following this, the papers published in 2006 and 2008 yielded 417 and 315 citations, respectively. The average number of citations a year was about 158, with a median of 129 and a standard deviation of 141.66.

As the result of a more in-depth study, it was evidenced that the first paper published in 2003 is called Cooperation Strategies for Agent-based P2P Systems \cite{penserini2003cooperation}. Following this, papers about ``Agents Handling Annotation Distribution'' were published in a corporate semantic Web \cite{gandon2003agents}, with another entitled: ``A new swarm mechanism based on social spider colonies: from web weaving to region detection'' \cite{bourjot2003new}. All three papers dealt with the WEB  AGENTS field and addressed areas involving multi-agent systems. Their order was determined based on the chronological timeline of the pages because they all arrived and were accepted on the same date, 21st March 2003.
 
\begin{table*}[!htb]
\centering
\label{artigos_mais_citados}
\caption{Top 10 most cited works.}
\renewcommand{\tabcolsep}{20 pt}
\resizebox{\linewidth}{!}{%
\begin{tabular}{rllrr}
\hline
\multicolumn{1}{c}{\textbf{Rank}} &
  \multicolumn{1}{c}{\textbf{Paper}} &
  \textbf{Year} &
  \multicolumn{1}{c}{\textbf{Cited in Scopus}} &
  \multicolumn{1}{c}{\textbf{Cited in Scholar}} \\ \hline
1  & Ontology-based personalized search and browsing                                              & 2003 & 304 & 686 \\
2  & Simplification and analysis of transitive trust networks                                     & 2006 & 152 & 258 \\
3  & Exploring local community structures in large networks                                       & 2008 & 122 & 310 \\
4  & WSMO-MX: A hybrid Semantic Web service matchmaker                                            & 2009 & 77  & 134 \\
5  & Social media gerontology: Understanding social media usage among older adults                & 2015 & 60  & 108 \\
6 &
  \begin{tabular}[c]{@{}l@{}}Combining multidimensional user models and knowledge representation and management \\ techniques for making web services knowledge-aware\end{tabular} &
  2006 &
  58 &
  58 \\
7  & Identifying document topics using the wikipedia category network                             & 2009 & 48  & 249 \\
8  & A web-based bayesian intelligent tutoring system for computer programming                    & 2006 & 44  & 124 \\
9  & Emergence of coordination in scale-free networks                                             & 2003 & 43  & 72  \\
10 & A new swarm mechanism based on social spiders colonies: From web weaving to region detection & 2003 & 39  & 87  \\ \hline
\end{tabular}%
}
\end{table*}

For this reason, Table \ref{artigos_mais_citados} outlines the most influential works based on the number of citations yielded in these two decades. The data show that the most commonly cited papers were published during the first five years of the Journal (2003-2008). These studies address questions ranging from ontology to trust networks and culminate in exploring the large networks \cite{gauch2003ontology, josang2006simplification, luo2008exploring} Only one of the ten most cited studies was published recently (2015). These works analyze subjects such as using social networks for the elderly \cite{hutto2015social}.Another critical fact is that the paper ``Combines multidimensional user models and knowledge representation and management techniques for making web services knowledge-aware'' \cite{cuzzocrea2006combining} is the only paper that has the same number of citations (58) in the Scopus database and Google Scholar (GS). In the case of the others, it is clear that the GS has a more significant number of citations for each paper. According to the findings of previous research, the reason for this is that GS detects documentary material in sources that do not belong to the journals, such as theses, books, the minutes of conferences, and unpublished material; this amounts to 50\% of all the material retrieved \cite{martin2018google}.

\begin{table}[!htb]
\centering
\label{top_instituicoes}
\caption{Top 10 Affiliations.}
\resizebox{\columnwidth}{!}{%
\begin{tabular}{rlr}
\hline
\multicolumn{1}{c}{\textbf{Rank}} & \multicolumn{1}{c}{\textbf{Affiliations}}                 & \multicolumn{1}{c}{\textbf{Documents}} \\ \hline
1  & Vrije Universiteit Amsterdam, Netherlands      & 10 \\
2  & University of Tokyo, Japan                     & 8  \\
3  & Chinese Academy of Science, China              & 7  \\
4                                 & CNRS Centre National de la Recherche Scientifique, France & 6                                      \\
5  & Queensland University of Technology, Austrália & 6  \\
6  & Wuhan University of Technology, China          & 6  \\
7  & Chongqing University, China                    & 6  \\
8  & University of Technology Sidney, Austrália     & 6  \\
9  & Ministry of Education China, China             & 5  \\
10 & University of Calgary, Canada                  & 5  \\ \hline
\end{tabular}%
}
\end{table}

\begin{figure*}[!htb] 
  \includegraphics[width=1\linewidth]{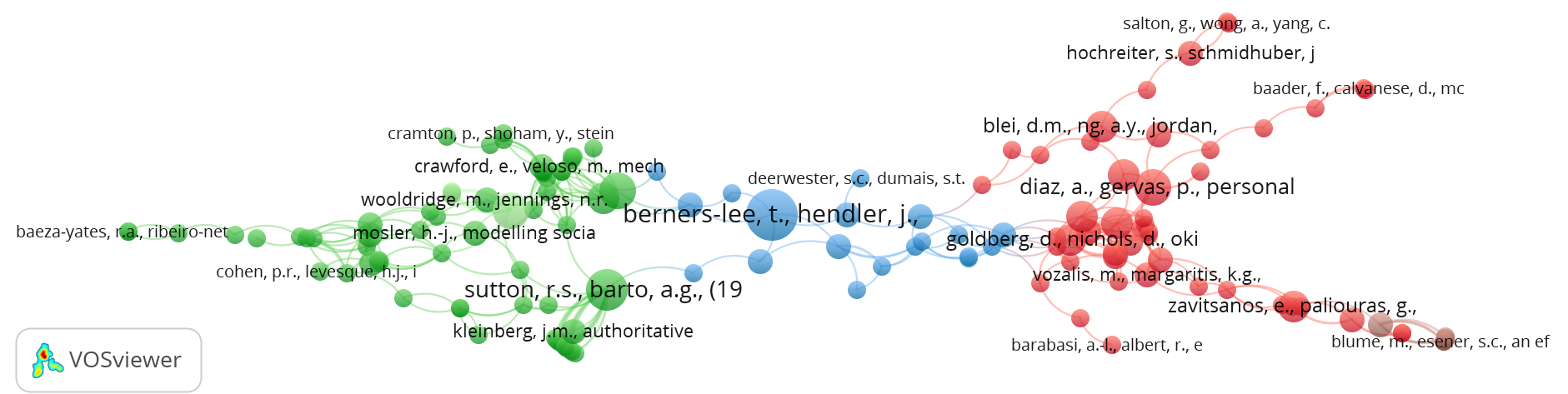}
  \caption{Co-citation analysis.}
  \label{co-citation}
\end{figure*}

Another strategy employed in this study was to analyze the references' co-citations and the published papers' keyword co-occurrence. This was to answer the third research question – namely, to determine the most popular topics that have emerged from the published works in the community. This approach makes it possible to establish the kind of relationship the studies have with the database regarding the number of times they are cited together \cite{small1973co}. Fig. \ref{co-citation} shows the co-citation analysis results using the cited references. It should be noted that three large groups could be identified: a) Blue – semantic and ontological; b) Green – multi-agent systems and intelligent systems; and c) Red - machine learning.

\subsection{ What is the current trend with regard to the publication and the most popular topics that have emerged from the  WIJ community?}\label{RQ3}

\begin{figure*}[!htb] 
  \includegraphics[width=.95\linewidth]{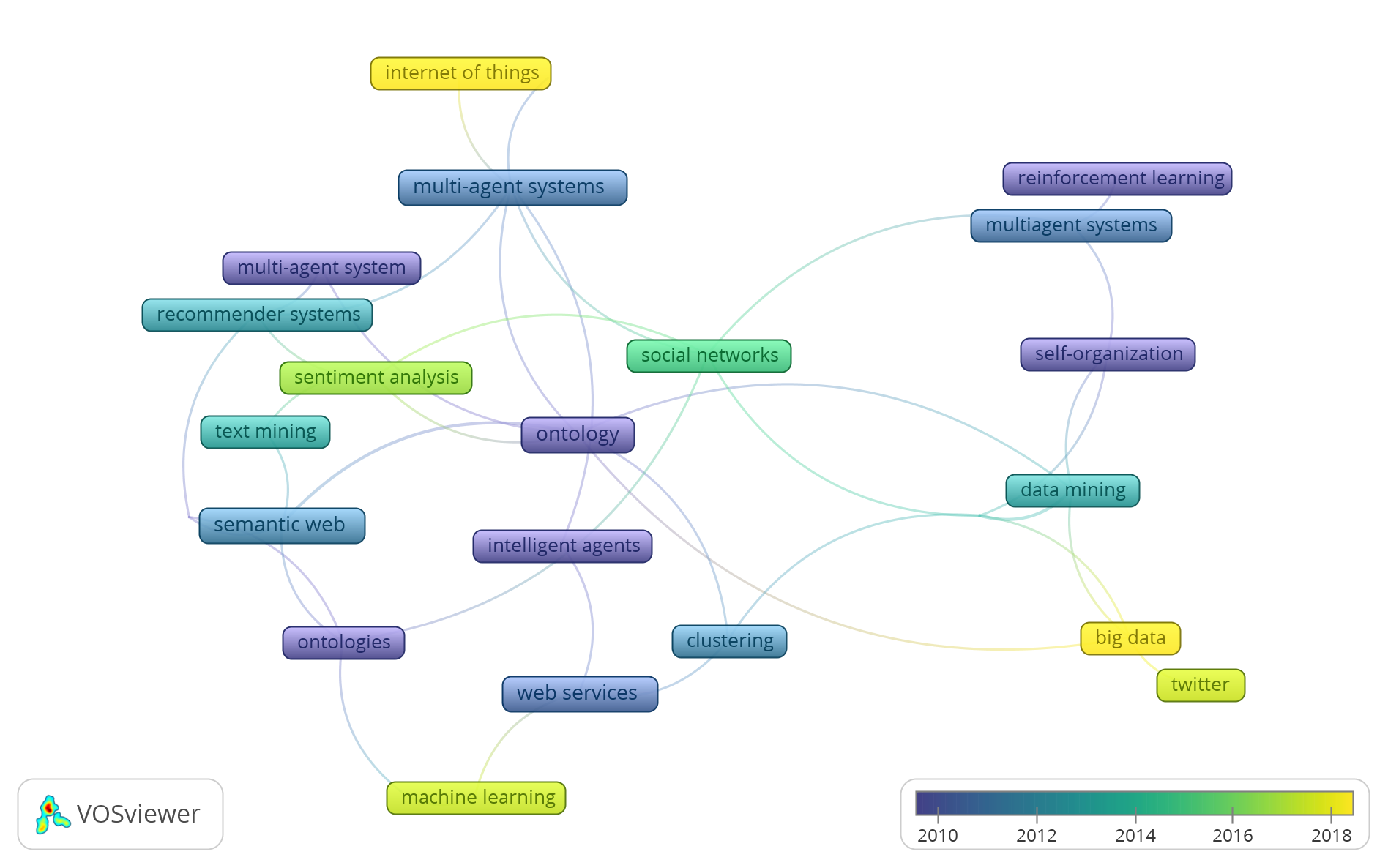}
  \caption{Author's keywords co-occurrence network over time}
  \label{fig:authors_keywords}
\end{figure*}

It is argued in the literature that the analysis of co-occurrences of the same word within a pair of words can indicate the main patterns, trends, and themes that can be found in a research field \cite{ding2001bibliometric, comerio2019tourism}. The  Tables \ref{tab:top_20_authors_keywords} and \ref{tab:top_20_index_keywords} show the keywords provided by the authors and indexed in the Scopus database, respectively, which have the largest number of occurrences in the publications of these two decades. It should be pointed out that what is indexed does not always reflect what the author enters. Furthermore,   VOSviewer Software \cite{van2010software} was employed to analyze the co-occurrence of the author's keywords by selecting keywords with a minimum of 5 occurrences. Fig. \ref{fig:authors_keywords} shows the relationship and the popularity of the authors' keywords when weighted by the average number of occurrences over time. It should be noted that the oldest studies are related to multi-agent systems, intelligent agents, and reinforcement learning, while the emerging issues are related to emotion analysis, machine learning, and big data, as well as the use of Twitter as a data source.

\begin{table}[!htb]
\caption{Top Author's keywords by frequency of their occurrence.}
\label{tab:top_20_authors_keywords}
\renewcommand{\tabcolsep}{5 pt}{%
\begin{tabular}{rlr}
\hline
\textbf{Rank} & \multicolumn{1}{c}{\textbf{Author's keywords}} & \textbf{Occurrences} \\ \hline
1             & multi-agent systems                            & 18                   \\
2             & ontology                                       & 15                   \\
3             & semantic web                                   & 14                   \\
4             & web services                                   & 10                   \\
5             & multiagent systems                             & 9                    \\
6             & social networks                                & 9                    \\
7             & big data                                       & 8                    \\
8             & cloud computing                                & 8                    \\
9             & machine learning                               & 8                    \\
10            & recommender systems                            & 8                    \\
11            & sentiment analysis                             & 8                    \\
12            & data mining                                    & 7                    \\
13            & reinforcement learning                         & 7                    \\
14            & clustering                                     & 6                    \\
15            & intelligent agents                             & 6                    \\
16            & multi-agent system                             & 6                    \\
17            & personalization                                & 6                    \\
18            & self-organization                              & 6                    \\
19            & information retrieval                          & 5                    \\
20            & internet of things                             & 5                    \\ \hline
\end{tabular}}%
\end{table}

\begin{table}[!htb]
\caption{Top Index keywords by frequency of their occurrence}
\label{tab:top_20_index_keywords}
\renewcommand{\tabcolsep}{5 pt}{%
\begin{tabular}{rlr}
\hline
\textbf{Rank} & \multicolumn{1}{c}{\textbf{Index keywords}} & \multicolumn{1}{c}{\textbf{Occurrences}} \\ \hline
1  & multi agent systems               & 73 \\
2  & semantics                         & 48 \\
3  & data mining                       & 43 \\
4  & intelligent agents                & 43 \\
5  & information retrieval             & 40 \\
6  & learning systems                  & 35 \\
7  & world wide web                    & 34 \\
8  & websites                          & 33 \\
9  & ontology                          & 32 \\
10 & social networking   (online)      & 32 \\
11 & algorithms                        & 30 \\
12 & semantic web                      & 30 \\
13 & search engines                    & 29 \\
14 & artificial   intelligence         & 27 \\
15 & problem solving                   & 26 \\
16 & distributed computer   systems    & 24 \\
17 & web services                      & 23 \\
18 & decision making                   & 22 \\
19 & mathematical models               & 21 \\
20 & classification (of   information) & 19 \\ \hline
\end{tabular}%
}
\end{table}

\begin{figure}[!htb] 
  \includegraphics[width=1\linewidth]{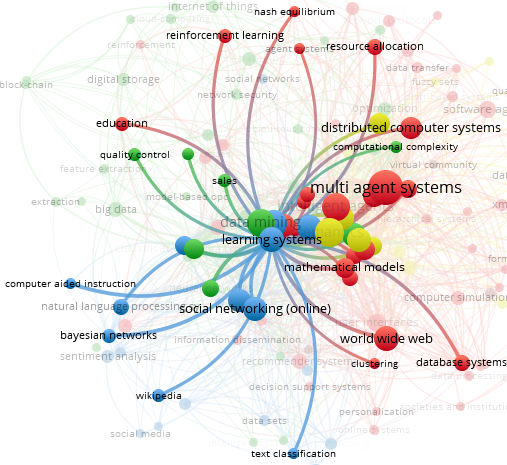}
  \caption{Learning system keywords network}
  \label{fig:key_learning_system}
\end{figure}

\begin{figure}[!htb] 
  \includegraphics[width=1\linewidth]{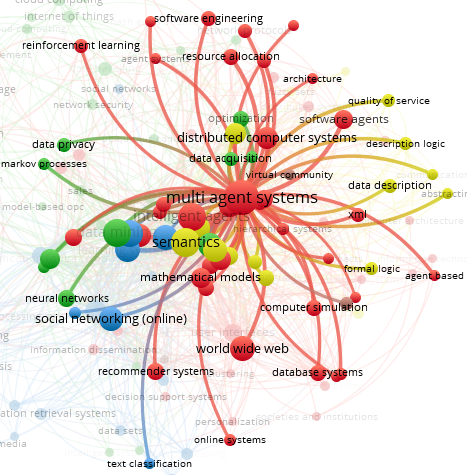}
  \caption{Multi agents system keywords network}
  \label{fig:key_multiagentes}
\end{figure}

\begin{figure}[!htb] 
  \includegraphics[width=1\linewidth]{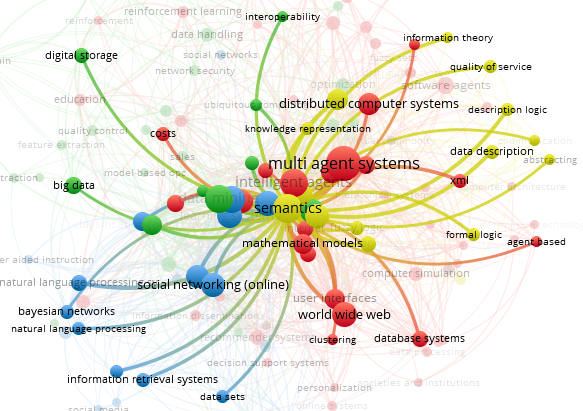}
  \caption{Semantics keywords network}
  \label{fig:key_semantics}
\end{figure}

\begin{figure}[!htb] 
  \includegraphics[width=1\linewidth]{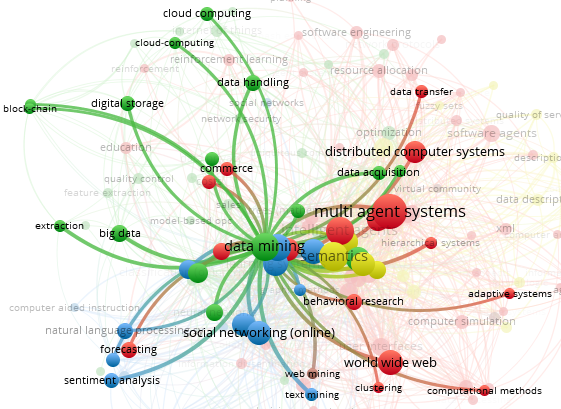}
  \caption{Data mining keywords network}
  \label{fig:key_datamining}
\end{figure}

\begin{figure*}[!htb] 
  \includegraphics[width=.95\linewidth]{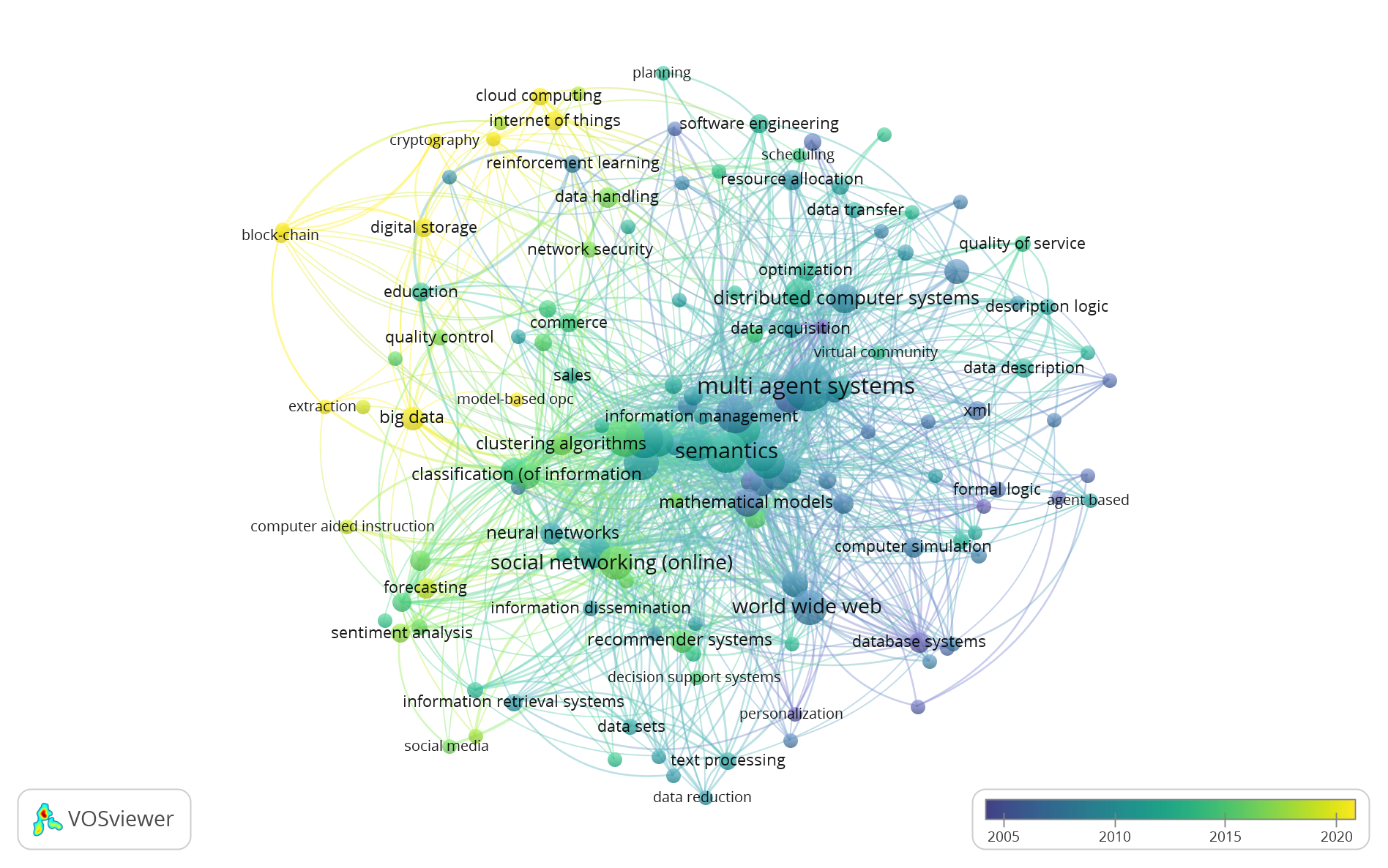}
  \caption{Index keywords co-occurrence network over time}
  \label{fig:index_keywords}
\end{figure*}

At the same time, Fig. \ref{fig:index_keywords} shows the relationship and popularity of the keywords indexed by the Scopus database when weighted by the average number of occurrences a year throughout the two decades. Following the pathway indicated by the authors' keywords, it is evident that the emerging issues concern machine learning, big data, ICT, cloud computing, and blockchain. The analysis of the relationship of the keywords enables four large topic clusters to be identified, which are related to the following: a) Blue – learning systems, b) Red – multi-agent systems, c) Yellow -  Semantics, and d) Green – data mining, as shown in  Figs. \ref{fig:key_learning_system}, \ref{fig:key_multiagentes}, \ref{fig:key_semantics}, \ref{fig:key_datamining}.

The emerging issues determined by the study meet the requirements of the  Gartner Hype Cycle Report 2022, which can be found in the decline of the curve of inflated expectations. This report provides a graphic representation of the maturity and adoption of technologies and how they evolve over time. The Hype Cycle can be used to identify an emerging technology in a particular sector \cite{HypeCycle:online}.

\section{Final Remarks}\label{FinalConsiderations}

This study analyzed the evolution of the Web Intelligence Journal by synthesizing the papers published in these two decades through a systematic review using scientometrics and bibliometrics approaches. The study can be distinguished from other literature reviews in so far as it does not focus on a specific field of knowledge but provides a general overview of the publications in WIJ over 20 years.

This study included a descriptive analysis of the published papers, the number of new authors each year, and the degree of inter-institutional cooperation. As a result, the way the community evolved over 20 years of publishing gradually became apparent. The study also showed which papers had the most influence on the community by analyzing the citation networks. The data showed that the most commonly cited papers were published in the opening years of WIJ. These studies address issues that refer to ontology, trust networks, and the exploration of large networks.

In addition, the authors analyzed the network in which links were forged between members of the community and the institutions to which they were affiliated, owing to the techno-scientific character of the publications. The analysis confirmed that the WIJ has global coverage and that China has the most institutions with publications, followed by the United States and Japan. However, Holland has the university with the most frequent number of publications.
 
Finally, the study showed the topics that were most popular and most often appeared among the published works in the community. This was achieved by employing a co-citation analysis of the references and the co-occurrence of keywords. This analysis evidenced three large groups: i) semantic and ontological, ii) multi-agent systems and intelligent agents, and iii) machine learning. In the same area, the analysis of the co-occurrence of the keywords indexed in the database allowed topics related to four large clusters to be identified: emotion analysis, machine learning, big data, ICT, cloud computing, and blockchain.

\section{Acknowledgments }\label{acknowledgments }

We are grateful to the National Council of Scientific Development and Technology  (CNPq) - DT-308334/2020; THE Amazon Foundation for Research Grants and Scholarships  (FAPESPA) - PRONEM-FAPESPA/CNPq nº 045/2021; and the Technical Cooperation Agreement nº 02/2021 (Processo n° 38328/2020 - TJ/MA), for partially funding this research project.

\bibliographystyle{unsrtnat}

\bibliography{thebibliography}

\end{document}